\documentclass[conference]{IEEEtran}
\usepackage{blindtext,amsmath,amssymb}
\usepackage{graphicx}
\usepackage{caption,subcaption}
\usepackage{tabu}
\usepackage{easytable}
\usepackage{array}

\newcolumntype{M}[1]{>{\centering\arraybackslash}m{#1}}
\newcolumntype{N}{@{}m{0pt}@{}}

\newcommand\norm[1]{\left\lVert#1\right\rVert}
\newcommand{\quotes}[1]{``#1''}
\ifCLASSINFOpdf
\else
\fi
\begin{document}
%
\title{EEG Signal Classification using Variational Mode Decomposition}

\author{\IEEEauthorblockN{Akshtih Ullal}
\IEEEauthorblockA{Department of EECS\\
Vanderbilt University\\
Email: akshith.ullal@vanderbilt.edu}
\and
\IEEEauthorblockN{Ram Bilas Pachori}
\IEEEauthorblockA{Department of EE\\
Indian Institute of Technology, Indore\\
Email: pachori@iiti.ac.in}}


%


\maketitle

\begin{abstract}
Epilepsy affects about 1\% of the population every year, and is characterized by abnormal and sudden hyper-synchronous excitation of the neurons in the brain. The electroencephalogram(EEG) is the most widely used method to record brain signals and diagnose epilepsy and seizure cases. In this paper we use the method of Variational Mode Decomposition (VMD) in our analysis to classify seizure/seizure free signals. This technique uses variational non recursive mode decomposition, in comparison to other methods like Empirical Mode (EMD) and Hilbert-Huang transform which recursively decompose the signals, making them more susceptible to noise and sampling rate. VMD decomposes a signal into its components which are called principal modes. In our analysis, 4 features of the decomposed signals namely  Renyi Entropy, second order difference plot (SODP), fourth order difference plot(FODP) and average amplitude are investigated, both individually and using a ranking methodology considering all 4 features at the same time.  The SODP of decomposed signal modes is an elliptical structure. The 95\% confidence ellipse area measured from the SODP of the decomposed signal modes has been used as a feature in order to discriminate seizure-free EEG signals from the epileptic seizure EEG signal. For the classification, a Multilayer Perceptron(MLP) with back propagation algorithm as the training method was used. A high percentage of accuracy was obtained when the features were used individually for classification and an even higher degree of accuracy was obtained when the ranking methodology was used.
\end{abstract}


%
\IEEEpeerreviewmaketitle

\section{Introduction}
Epilepsy is the fourth most common neurological disorder and affects people of all ages.[1] Epilepsy is one of the reasons seizures are caused, but may not be the only reason that seizures occur[2]. During the seizure phase there is a lot of uncontrolled electrical disturbances taking place between the neurons of the brain. The electroencephalogram(EEG) signal is the most common way to measure neural activity on the surface of the brain, and is widely used in the detection of epilepsy[3]. Generally normal activity of the brain consists of non-synchronous signals, as the neurons are firing in different directions, when the brain is performing different activities[4]. However, during seizure the electrical activity can become highly synchronous which is what we are interested in looking at. Not all seizures are caused by epilepsy. Some other causes include psychogenic non-epileptic seizures, syncope, sub-cortical movement disorders and migraine variants[5]. Our main goal in this paper is to be able to classify epileptic from non-epileptic seizures.  
Although EEG signals can be noisy and contain artefacts, generally they can be classified as having 5 frequency bands, namely delta, theta, alpha, beta and gamma bands[6]. In the earlier days, EEG signal analysis was conducted by assuming the signal to be stationary. Hence time and frequency analysis were conducted independently as in [7][8]. The problem with this approach was that, a time domain analysis would lose resolution in frequency and vice versa, which meant some information would be lost. In order to tackle this problem, the time-frequency domain analysis was used in [9][10] by considering the EEG signal to be non-stationary. These techniques included using wavelet and multi-wavelet transforms[11-17]. In addition to time-frequency methods, others include parametric, non-parametric and eigen-vector methods. \par
The parametric model, as the name suggests is based on the assumption that the EEG signal satisfies a particular generating model and its behavior can be described by using a formula with a number of parameters to be estimated. Some of the widely used parametric methods are Auto-regressive model, Moving Average model, Auto-regressive moving average model[18]. The AR model is very good at representing signals with high and narrow peaks. The Moving Average model is better at representing broad peaks. Non-parametric methods rely on calculating the power spectral density(PSD) in order to represent spectral resolution. There are 2 main types which are Periodogram and Correlogram. These 2 techniques provide reasonable resolution for sufficient data lengths. However they  suffer with the problem of having a high variance, even for longer data lengths and methods have been tried in [19] to reduce the variance by making a compromise on the resolution.  Eigen vector methods are used for estimating frequencies when the signals have been corrupted by noise[20].
The techniques of wavelet transform are not exactly completely time-frequency analysis, as they divide the signal into small length of time, but lose resolution within those small segments. Recently signal decomposition techniques like Empirical Mode Decomposition(EMD) [21]  have been used in EEG signal analysis for analysing non-linear and non-stationary signals and many of these techniques have been applied to EEG signals[22-28]. The idea behind EMD is to decompose the original signals into its Intrinsic Mode Functions(IMF). IMFs are derived signal components which have inherent characteristics of the original signals. They mainly satisfy 2 conditions. 1) In the whole signal, the number of extrema and the number of zero crossings must be the same or at most differ by one 2) At any point, the mean value of the envelope defined by the local maxima and the envelope defined by the local minima should be zero[29]. In other words, these conditions define the IMFs as AM-FM signals. With some shortcomings,  EMD  algorithm has been fairly widely used in many applications such as, signal decomposition in audio engineering [30], climate analysis [31], and various flux, respiratory, and neuromuscular signals found in medicine and biology [32-35]. The problem with EMD is that there is a lack of mathematical theory behind it as the algorithm is empirical in nature, Hence the algorithm's results are highly dependent on the methods of extremal point finding, interpolation of extremal points into carrier envelopes, and the stopping criteria imposed, which are techniques that are subjective in nature [29]. 
In order to overcome some of the drawbacks of EMD, a new technique was developed called Variational Mode Decomposition (VMD) in [29]. One of the main drawbacks of EMD is that it is recursive,  which does not allow backward error correction[36]. In contrast, VMD is an intrinsic, and variational method that determines the frequency band of each mode adaptively and the modes concurrently. Hence if there is an error in one of the iterations while calculating the frequency bands, it can be quickly balanced while calculating the modes.  The noise characteristics obtained during VMD has a close relation to Wiener filtering [29]. Our main aim in this paper is to apply the VMD algorithm in order to classify EEG signals.

\section{Methodology}
\subsection{Dataset}
The EEG dataset used in this analysis is available publicly online[37] and has been taken from the University of Boon website. The entire recorded data is divided into five groups namely Z, O, N, F and S which contain the signals of both healthy and seizure affected subjects. Each of this set has got 100 single channel EEG signal recording, that were obtained by sampling at a rate of 173.61Hz for a duration of 23.6 seconds. The sub-sets Z and O have been recorded using surface EEG recordings of five healthy volunteers with eyes open and closed, respectively, using standard electrode placement scheme according to the international 10-20 system. The signals in two subsets(N and F) have been recorded in seizure-free intervals from five patients in the epileptogenic zone (subset F) and from the hippocampal formation of the opposite hemisphere of the brain. The subset S includes seizure activity selected from all recording sites exhibiting ictal activity. The subsets N, F, and S have been recorded intracranially. All EEG signals were recorded with the same 128-channel amplifier system with an average common reference.  The main goal was to  check the accuracy of the proposed method in classifying between Normal/Seizure, Seizure/Seizure-free and Seizure/Non-Seizure signals. For the Normal/Seizure classification, the datasets Z, O, and S were used. For the Seizure/Seizure-free classification the datasets N, C and S were used and finally for the Seizure/Non-Seizure classification, all the five datasets of Z, O, N, C, and S were used.   
\subsection{Variational Mode Decomposition}
VMD uses variational techniques and the main objective is to decompose the original signal into its discrete number of principal modes or sub modes that have specific sparse properties, while also keeping in mind the reconstruction of the original signal from these modes. In the VMD algorithm we have chosen the bandwidth as the sparse property for each decomposed mode[29].Hence our final objective is to decompose the signal into its respective modes and also calculate their centre frequency \(u_k\).This means each of the sub-modes will have a centre frequency \(\omega_k\) which will be calculated during the decomposition. The calculation of bandwidth(centre frequency) for each mode is done as follows: 1) the analytical signal of the mode is calculated by taking the Hilbert transform of the signal which gives us the unilateral frequency spectrum of the mode. 2) Each mode signal is then shifted to baseband by mixing it with another signal which is tuned to the respective estimated centre frequency. 3) The bandwidth is now estimated by taking the squared L-2 norm of the gradient. The variational problem with the reconstruction constraint can be written as

 \begin{equation}
  min_{u_k,\omega_k}
  \bigg \{ \sum_{k}^{} \norm{\partial_t\bigg[\bigg(\delta\big(t\big)+\frac{j}{\pi t}\bigg)*u_k\big(t\big)\bigg]e^{-j\omega_k t}}_2^2\bigg \}
 \end{equation}
 
 The reconstruction constraint can be enforced in different ways. In this algorithm 2 methods are used namely, the Lagrange multiplier and the quadratic penalty method. The Lagrangian multiplier is used for enforcing the reconstruction constraint and the quadratic penalty method is used to increase the convergence of the result.  Applying this condition the above equation can be rewritten as follows
 
  \begin{equation}
  \begin{aligned}
 \mathcal{L}\big(u_k,\omega_k,\lambda\big)=\alpha \sum_{k}^{} \norm{\partial_t\bigg[\bigg(\delta\big(t\big)+\frac{j}{\pi t}\bigg)*u_k\big(t\big)\bigg]e^{-j\omega_k t}}_2^2 \\
 +\norm{f-\sum{u_k}}+\Big \langle \lambda,f-\sum{u_k} \Big \rangle .
 \end{aligned}
 \end{equation}
 
 To find the optimal centre frequency we need to minimize the above Lagrangian, in other words to find the saddle point. This is done in iterations using the alternate direction method of multipliers(ADMM) as shown in [29]. the 3 main equations of the iteration are shown below:

\begin{equation}
\hat{u}_k^n+1\leftarrow\frac{\hat{f}-\sum_{i<k}\hat{u}_i^{n+1}-\sum_{i<k}\hat{u}_i^n+\frac{\hat{\lambda}^n}{2}}{1+2\alpha\big(\omega-\omega_k^n\big)^2}
\end{equation}
\begin{equation}
\omega_k^{n+1}\leftarrow\frac{\int_{0}^{\infty}\omega\big|\hat{u}_k^{n+1}\big(\omega\big)\big|^2d\omega}{\int_{0}^{\infty}\big|\hat{u}_k^{n+1}\big(\omega\big)\big|^2d\omega}
\end{equation}
\begin{equation}
\hat{\lambda}^{n+1}\leftarrow\hat{\lambda}^n+\tau\Bigg(\hat{f}-\sum_{k}\hat{u}_k^{n+1}\Bigg)
\end{equation}
The theory of how the results of equation (4) (5)  are obtained is also shown in [29]
\begin{figure*}

\begin{minipage}{\textwidth}
 \centering
  \includegraphics[width=\textwidth,height=7.5cm]{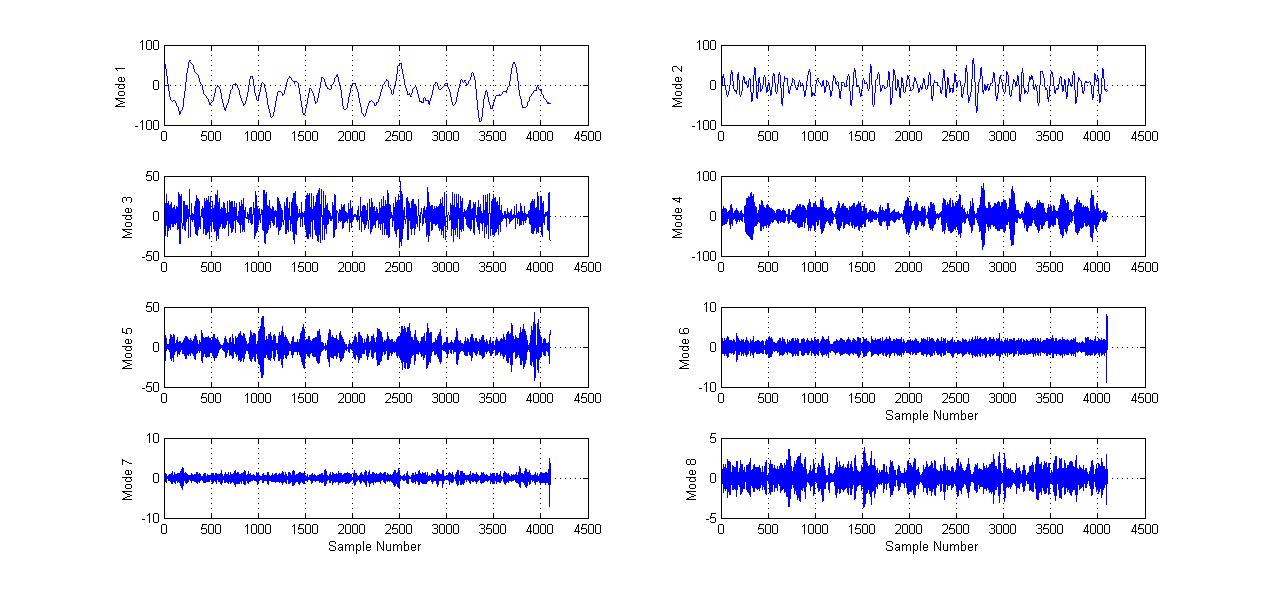}
  \subcaption{Variational Mode Decomposition of the 23.6s seizure free EEG signal}
  \includegraphics[width=\textwidth,height=7.5cm]{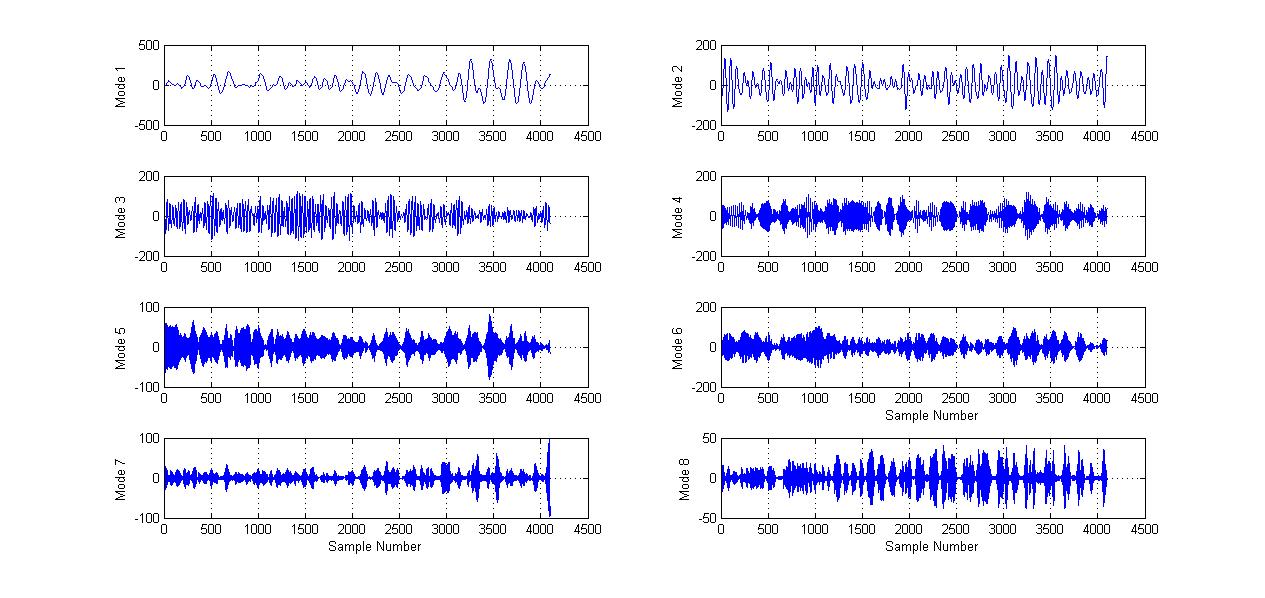}
  \subcaption{Variational Mode Decomposition of the 23.6s ictal EEG signal}
 \end{minipage}
\caption{Comparison of seizure-free and ictal signals after Variaontional Mode Decomposition. Ictal signals have a much larger amplitude compared to seizure-free signals across the different modes \label{overflow}}
\end{figure*}
\begin{figure*}
\begin{minipage}{\textwidth}
 \centering
\includegraphics[width=\textwidth,height=7.5cm]{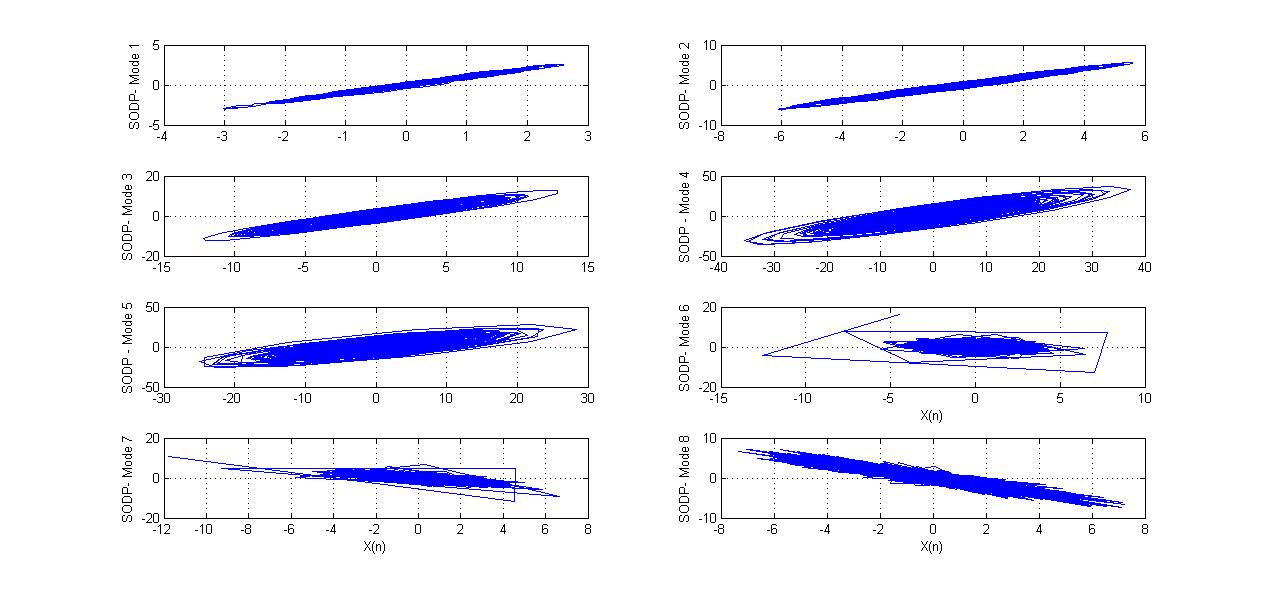}
  \subcaption{SODP of the principal modes of a  seizure free EEG signal}
\includegraphics[width=\textwidth,height=7.5cm]{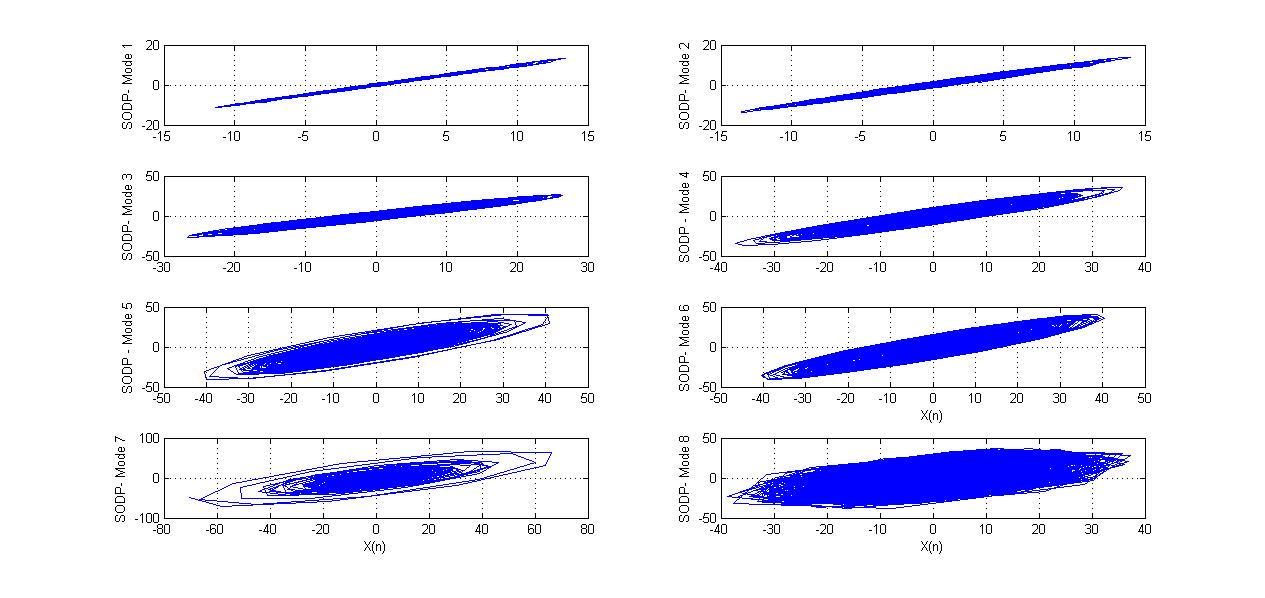}
  \subcaption{SODP of the principal modes of an ictal EEG signal}
 \end{minipage}
\caption{Comparison of the SODPs of seizure-free and ictal signals.It can be seen that seizure free SODPs tend to change the orientation of their ellipse at higher modes.   }
\end{figure*}
\begin{figure*}
\begin{minipage}{\textwidth}
 \centering
\includegraphics[width=\textwidth,height=7.5cm]{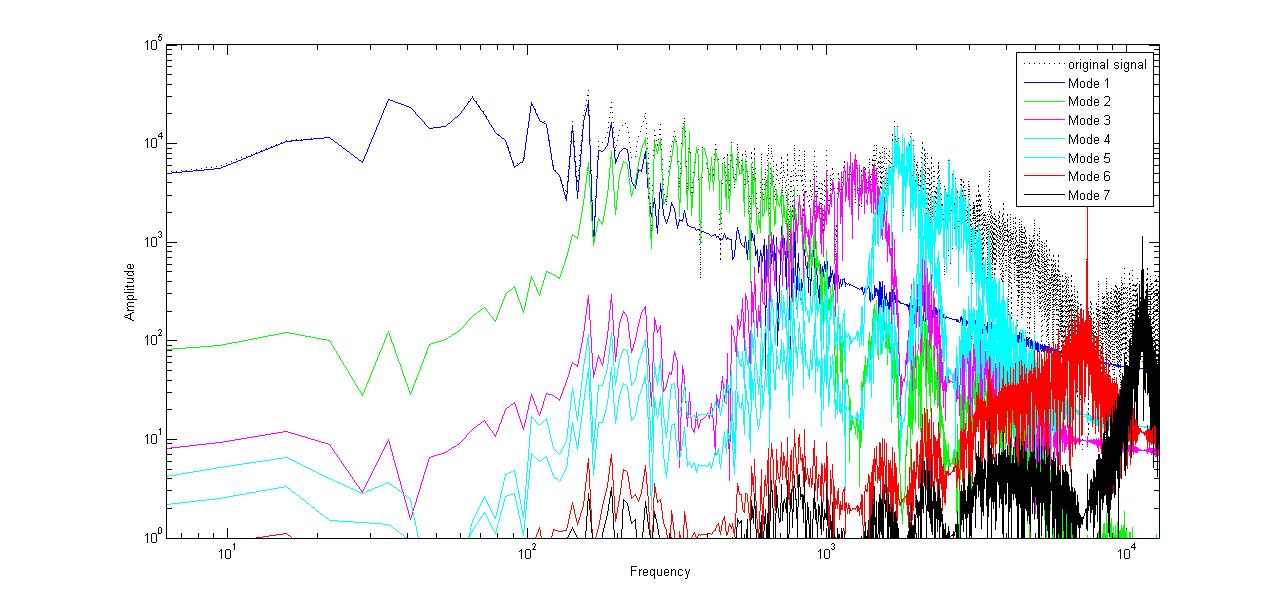}
  \subcaption{Frequency spectrum of the principal modes of a seizure free signal}
\includegraphics[width=\textwidth,height=7.5cm]{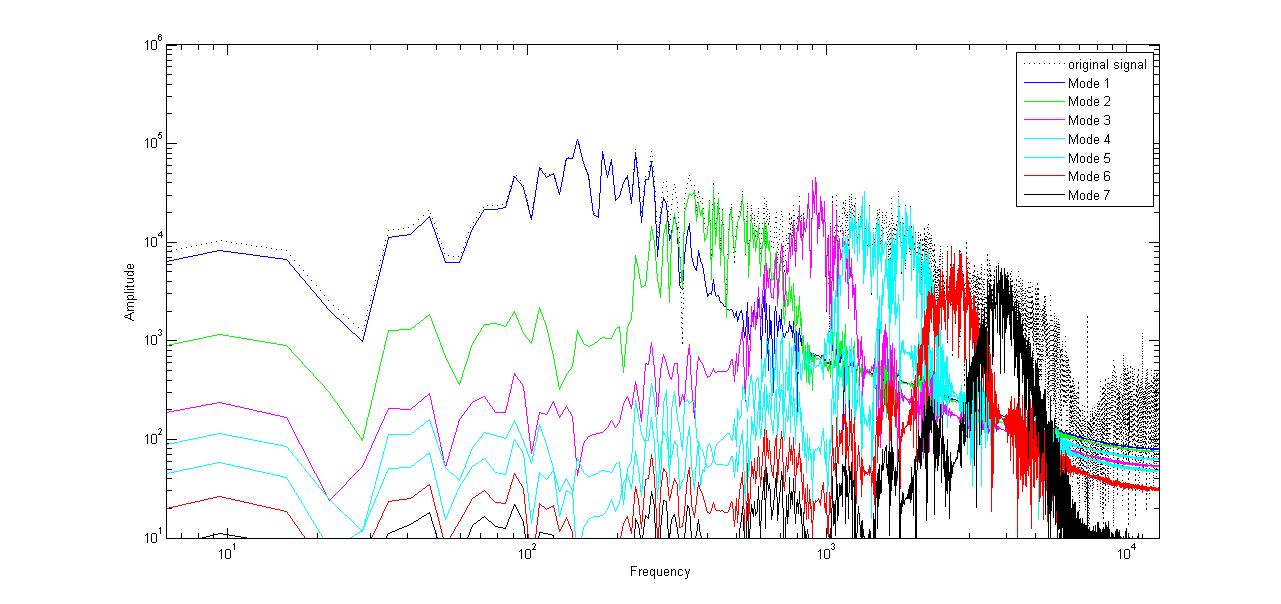}
  \subcaption{Frequency spectrum of the principal modes of an ictal signal}
 \end{minipage}
\caption{Comparison of the spectrum of seizure-free and ictal signals.}
\end{figure*}
\subsection{Second and Fourth order difference plot and computation of ellipse area}
The second order difference plot (SODP)[39]  of the decomposed VMD modes can be a helpful tool in classifying seizure and non-seizure signals. The SODP intuitively tells us the rate of variations of successive samples of the signal. Suppose we have a signal x(n), its SODP can be calculated by plotting X(n) against Y(n) which are defined as [38]

\begin{equation}
X\big(n\big)=x\big(n+1\big)-x\big(n\big)
\end{equation}
\begin{equation}
Y\big(n\big)=x\big(n+2\big)-x\big(n+1\big)
\end{equation}

Recently, SODP has been used in variability analysis for EEG signals and Centre of pressure(COP) signals. The 95\% confidence area has been used to calculate the SODP area. The area is calculated for each of the decomposed signal mode. The method used to calculate the SOPD area is given below:

Compute the mean values of X(n) and Y(n)
\begin{equation}
S_X=\sqrt{\frac{1}{N}\sum_{n=0}^{N-1}X\big(n\big)^2}
\end{equation}
\begin{equation}
S_Y=\sqrt{\frac{1}{N}\sum_{n=0}^{N-1}Y\big(n\big)^2}
\end{equation}
\begin{equation}
S_{XY}=\frac{1}{N}\sum X\big(n\big)Y\big(n\big)
\end{equation}

Compute the D parameter as:
\begin{equation}
D=\sqrt{\big(S_X^2+S_Y^2\big)-4\big(S_X^2S_Y^2-S_{XY}^2\big)}
\end{equation}

\begin{equation}
a=1.7321\sqrt{S_X^2+S_Y^2+D}
\end{equation}
\begin{equation}
b=1.7321\sqrt{S_X^2+S_Y^2-D}
\end{equation}
From the parameters a and b, the ellipse area is calculated as: 
\begin{equation}
A_{ellipse}=\pi ab
\end{equation}
For the fourth order difference plot(FODP),the procedure is same as SODP except that we shift the signal sample by 2 samples in both the X and Y axis. Hence the X(n) and Y(n) in equations (7) and (8) change to

\begin{equation}
X\big(n\big)=x\big(n+2\big)-x\big(n\big)
\end{equation}
\begin{equation}
Y\big(n\big)=x\big(n+4\big)-x\big(n+2\big)
\end{equation}
The rest of  process remains the same as in SOPD calculations.

\subsection{Renyi Enntropy}
The idea of using entropy to classify signals stems from the fact that entropy characterises  the distortedness or randomness of the signal. Generally normal signals tend to have more random impulses compared to ictal signals. Spectral entropy [40] is evaluated using the normalized Shannon entropy, which quantifies the spectral complexity of the time series. Fourier transformation is used to obtain the power spectral density (PSD) of the time series. The PSD represents the distribution of power of the signal according to the frequencies present in the signal. In order to obtain the power level for each frequency, the Fourier transform of the signal is computed, and the power level of the frequency component is denoted by $P_f$ . The normalization of the power is performed by computing the total power as $\sum P_f$ and dividing the power level corresponding to each frequency by the total power as:
\begin{equation}
p_f=\frac{P_f}{\sum P_f}
\end{equation}
We have used the Renyi entropy as the feature to classify EEG signals[41], which can be generally defined as:
\begin{equation}
RenEn\big(\alpha\big)=\frac{1}{1-\alpha}log\Bigg(\sum_{f} p_f^\alpha\Bigg) ,\alpha>0
\end{equation}
In our case we use $\alpha$ =2 which is also known as quadratic Renyi entropy.
\subsection{Average Amplitude}
Using the VMD algorithm to decompose the modes of signal, the average value of the amplitude was one of the simple yet distinguishable characters used in the classification. It is calculated by taking the sum of the absolute value of all

the samples and dividing it by 4097, i.e the total number of samples.
\subsection{Multilayer Perceptron (MLP) based classification}
The classifier used for this analysis is the Multilayer Perceptron(MLP) which is a feed forward artificial neural network (ANN). ANNs are inspired to emulate the working of the human brain and contain different number of nodes which can be distributed across many layers, with all the elements of one layer connected to the next one. Each layer has got an activation function and a weight that connects it to another node of the next layer. In our classifier we used a sigmoid activation function for each node. The configuration that yielded the best results were 2 hidden layers, each consisting of 10 nodes each and an input and output layer.

\section{Results}
Variable Mode Decomposition(VMD) decomposes a given signal into its principal modes. The analysis was done using 4 features extracted from the principal nodes which are, the Renyi entropy, the second order difference plot(SODP), fourth order difference plot(FODP) and average amplitude. The features were tested individually and also combining all the features, i.e all the above features were used for classification with a ranking methodology used between them. In VMD, the first principal mode has the lowest frequency. The consecutive modes constitute the higher frequencies as seen in Fig 1, unlike in Empirical Mode Decomposition(EMD) where the first few modes, constitute the higher frequencies. Among the 4 features individually investigated, The Renyi entropy has the least accuracy, the results of which are shown in Table 1. This table shows the results for which the first 80 samples were used as the training data, and the remaining 20 samples(each set(Z,O,N,F,S) contains 100 samples) were used as classification test data.  The entropy was calculated for each of the decomposed modes and fed as separate inputs to the MLP classifier. The results for Normal/Seizure classification (Z,O and S) are higher, as this set consists of signals from healthy patients and confirmed seizure patients. The Seizure/Seizure-free(N,F and S) and Seizure/Non-Seizure(Z,O,N,F and S) classifications consist of the N and F datasets which are seizure free, but may have other brain conditions, which increases the risk of misclassification and reduces the accuracy. \par
The SODP of the decomposed modes is plotted in Fig 2. We can observe the elliptical structure of the modes. Hence we can calculate the area of the ellipse. We have used 95\% confidence ellipse area as a feature for discriminating ictal EEG signals from the seizure-free EEG signals. The results obtained are shown in Table 1. The second order difference plot concept can be extended to the fourth-order by delaying the signal samples twice the amount. This is called the fourth order difference plot (FODP). The calculation of the ellipse area remains the same as in the case of SODP.  The final feature used is the average amplitude of each of the decomposed modes. Compared to EMD, the difference in amplitude using VMD is more pronounced between the decomposed modes of seizure free and ictal signals which gives a high degree of accuracy. The results are again shown in Table 1.

\begin{table}[ht]
\caption {Accuracy(\%) with using features individually} \label{tab:title}
\begin{tabular}{|m{2.2cm}|m{1.5cm}|m{1.5cm}|m{1.5cm}|N} 
\hline
Feature/Accuracy(\%) & \textbf{Normal/Seizure} & \textbf{Seizure/Seizure-free} & \textbf{Seizure/Non-Seizure} &\\[10pt]
\hline
\textbf{SODP} & 93.3 & 96.6 & 96 &\\[10pt] 
\hline
\textbf{FODP} & 91.6 & 100 & 96  &\\[10pt] 
\hline
\textbf{Average Amplitude} & 98.3 & 98.3 & 98 &\\[10pt] 
\hline
\textbf{Renyi Entropy} & 98.3 & 88.3 & 90 &\\[10pt] 
\hline
\end{tabular}

\end{table}
\begin{table}[ht]
\caption {Accuracy(\%) with features ranked} \label{tab:title}
\begin{tabular}{|m{2.2cm}|m{1.5cm}|m{1.5cm}|m{1.5cm}|N} 
\hline
Feature/Accuracy(\%) & \textbf{Normal/Seizure} & \textbf{Seizure/Seizure-free} & \textbf{Seizure/Non-Seizure} &\\[10pt]
\hline
\textbf{3 features} (except entropy) & 98.3 & 100 & 99 &\\[10pt] 
\hline

\end{tabular}

\end{table}

\par We have also tested using all 3 features in the classification logic. Renyi entropy was left out due to its lower accuracy. The ranking methodology used for this is as follows: The classification is initially done using each feature individually. Once the classification results for all 3 features are obtained, they are compared. If all three features output the same result, then that result will be the output. If the results differ, then the majority classification is taken as the result (Table II). In order to get rid of the bias of selecting a particular set of training and testing samples, for this classification 80 samples were randomly selected without replacement, and the remaining 20 samples which were not in the training set were used as the test set. Same feature ranking method was used as the previous method. 24 iterations were run, where in each iteration the training and test set is random. The results are shown below:

\begin{table}[ht]
\caption {Accuracy(\%) with training and testing data randomized(mutually exclusive)} \label{tab:title}
\begin{tabular}{|m{2.2cm}|m{1.5cm}|m{1.5cm}|m{1.5cm}|N} 
\hline
\textbf{Iteration} & \textbf{Normal/Seizure} & \textbf{Seizure/Seizure-free} & \textbf{Seizure/Non-Seizure} &\\[5pt]
\hline
1 & 98.3 & 96.3 & 96 &\\[1.5pt] 
\hline
2 & 100 & 98.3 & 97 &\\[2pt] 
\hline
3 & 100 & 95 & 98 &\\[2pt] 
\hline
4 & 98.3 & 98.3 & 97 &\\[2pt] 
\hline
5 & 98.3 & 96.3 & 99 &\\[2pt] 
\hline
6 & 96.3 & 96.3 & 97 &\\[2pt] 
\hline
7 & 100 & 98.3 & 99 &\\[2pt] 
\hline
8 & 100 & 98.3 & 97 &\\[2pt] 
\hline
9 & 100 & 95 & 100 &\\[2pt] 
\hline
10 & 96.3 & 93.6 & 98 &\\[2pt] 
\hline
11 & 100 & 93.6 & 98 &\\[2pt] 
\hline
12 & 100 & 100 & 95 &\\[2pt] 
\hline
13 & 96.3 & 93.6 & 99 &\\[2pt] 
\hline
14 & 95 & 98.3 & 98 &\\[2pt] 
\hline
15 & 98.3 & 100 & 98 &\\[2pt] 
\hline
16 & 98.3 & 96.3 & 98 &\\[2pt] 
\hline
17 & 96.3 & 93.6 & 98 &\\[2pt] 
\hline
18 & 95 & 96.3 & 97 &\\[2pt] 
\hline
19 & 96.3 & 96.3 & 100 &\\[2pt] 
\hline
20 & 98.3 & 98.6 & 95 &\\[2pt] 
\hline
21 & 96.3 & 93.6 & 96 &\\[2pt] 
\hline
22 & 100 & 93.6 & 99 &\\[2pt] 
\hline
23 & 98.3 & 95 & 100 &\\[2pt] 
\hline
24 & 98.3 & 98.6 & 100 &\\[2pt] 
\hline
\textbf{Average} & 98.2 & 96.4 & 97.9 &\\[2pt] 
\hline

\end{tabular}

\end{table}

\section{Conclusion}
In this paper, the Variational Mode Decomposition(VMD) method was applied to classify EEG signals as seizure-free/seizure. The VMD had many improvements over previous methods like EMD and Wavelet transforms. It has also given us promising results with an average accuracy of 98.2\% ,96.4\% and 97.9\% for the classification of Normal/Seizure, Seizure/Seizure-free and Seizure/Non-Seizure cases respectively.It is a slight improvement on the accuracy of 97.7\% obtained in [39]. and a significant improvement over the accuracy of 94\% and 95\% obtained in [7] and [8] respectively. However we would like to try our algorithm over a larger data set, as the Bonn data set in total has only 300-500 samples, based on the type of ,for both testing and training. Hence a single anomaly or misclassification would bring down the accuracy drastically. We are also looking at applying new features like Lypanouv exponent, Shanon entropy etc in our future work.


%



\ifCLASSOPTIONcaptionsoff
  \newpage
\fi




\begin{thebibliography}{1}
\bibitem{IEEEhowto:kopka}
	F. Mormann, R.G. Andrzejak, C.E. Elger, K. Lehnertz , \emph{Seizure prediction: the long and winding road},\hskip 1em plus
  0.5em minus 0.4em\relax vol 2, pp. 314-333, December 2007.
  \bibitem{IEEEhowto:kopka}
	L. Guerra, M. Moreno, J Chozas, R. Bolanos ,\quotes{Electrolyte Disturbance and Seizures},  \emph{Epilepsia The official journal of the International League against Epilepsy},\hskip 1em plus
  0.5em minus 0.4em\relax 28th November 2006.
\bibitem{IEEEhowto:kopka}
	L.D. Iasemidis, et al.,\quotes{Adaptive epileptic seizure prediction system},  \emph{IEEE Transactions on Biomedical Engineering},\hskip 1em plus
  0.5em minus 0.4em\relax pp. 616-627, May 2003.
  \bibitem{IEEEhowto:kopka}
		Hammer, edited by Stephen J. McPhee, Gary D ,  \emph{Pathophysiology of disease : an introduction to clinical medicine },\hskip 1em plus
  0.5em minus 0.4em\relax New York: McGraw-Hill Medical, 2010.
  \bibitem{IEEEhowto:kopka}
		Ann Pietrangelo, \relax (2017, January 10). [Online]. Available: http://www.healthline.com/health/epilepsy.
  \bibitem{IEEEhowto:kopka}
		Tatum, William O ,\quotes{Ellen R. Grass Lecture: Extraordinary EEG},  \emph{Neuro diagnostic Journal 54.1},\hskip 1em plus
  0.5em minus 0.4em\relax pp. 3-21, 2014.
  \bibitem{IEEEhowto:kopka}
		S. Altunay, Z. Telatar, O. Erogul ,\quotes{Epileptic EEG detection using the linear prediction error energy},  \emph{Expert System with Applications},\hskip 1em plus
  0.5em minus 0.4em\relax pp. 5661-5665, August 2010.
  \bibitem{IEEEhowto:kopka}
	V. Joshi, R.B. Pachori, A. Vijesh ,\quotes{Classification of ictal and seizure-free EEG signals using fractional linear prediction},  \emph{Biomedical Signal Processing and Control},\hskip 1em plus
  0.5em minus 0.4em\relax pp. 1-5, January 2014.
  \bibitem{IEEEhowto:kopka}
		A.T. Tzallas, M.G. Tsipouras, D.I. Fotiadis ,\quotes{Automatic seizure detection based on time-frequency analysis and artificial neural networks},  \emph{Computational Intelligence and Neuroscience 2007},\hskip 1em plus
  0.5em minus 0.4em\relax Article ID 80510, 2007.
  \bibitem{IEEEhowto:kopka}
		A.T. Tzallas, M.G. Tsipouras, D.I. Fotiadis ,\quotes{Epileptic seizure detection in EEGs using time-frequency analysis},  \emph{IEEE Transactions on Information Technology in Biomedicine },\hskip 1em plus
  0.5em minus 0.4em\relax pp. 703-710, September 2009.
  \bibitem{IEEEhowto:kopka}
	H. Adeli, Z. Zhou, N. Dadmehr ,\quotes{Analysis of EEG records in an epileptic patient using wavelet transform},  \emph{Epilepsia Journal of Neuroscience Methods},\hskip 1em plus
  0.5em minus 0.4em\relax pp. 69-87, February 2003.
  \bibitem{IEEEhowto:kopka}
	Y.U. Khan, J. Gotman,\quotes{Wavelet based automatic seizure detection in intracerebral electroencephalogram},  \emph{Clinical Neurophysiology},\hskip 1em plus
  0.5em minus 0.4em\relax pp. 898-908, May 2003.
  \bibitem{IEEEhowto:kopka}
	S. Ghosh-Dastidar, H. Adeli, N. Dadmehr,\quotes{Mixed-band wavelet-chaos-neural network methodology for epilepsy and epileptic seizure detection},  \emph{IEEE Transactions on Biomedical Engineering},\hskip 1em plus
  0.5em minus 0.4em\relax pp. 1545-1551, September 2007.
  \bibitem{IEEEhowto:kopka}
		H. Ocak ,\quotes{Automatic detection of epileptic seizures in EEG using discrete wavelet transform and approximate entropy},  \emph{Expert Systems with Applications },\hskip 1em plus
  0.5em minus 0.4em\relax pp. 2027-2036, March 2009.
  \bibitem{IEEEhowto:kopka}
	H. Adeli, S. Ghosh-Dastidar, N. Dadmehr,\quotes{A wavelet-chaos methodology for analysis of EEGs and EEG subbands todetect seizure and epilepsy},  \emph{IEEE Transactions on Biomedical Engineering },\hskip 1em plus
  0.5em minus 0.4em\relax pp. 205-211, February 2007.
  \bibitem{IEEEhowto:kopka}
		A. Subasi,\quotes{EEG signal classification using wavelet feature  extraction and a mixture of expert model},  \emph{Expert Systems with Applications },\hskip 1em plus
  0.5em minus 0.4em\relax pp. 1084-1093, May 2007.
  \bibitem{IEEEhowto:kopka}
	R. Uthayakumar, D. Easwaramoorthy ,\quotes{Epileptic seizure detection in EEG signals using multifractal analysis and wavelet transform}, \emph{Fractals 21 },\hskip 1em plus
  0.5em minus 0.4em\relax June 2013.
  \bibitem{IEEEhowto:kopka}
	Guler, I. And Ubeyli, E.D ,\quotes{Multiclass support vector machines for EEG signal Classification},  \emph{IEEE Transactions on Information Technology in Biomedicine},\hskip 1em plus
  0.5em minus 0.4em\relax Vol. 11, no. 2, pp.  117-126, 2007..
  \bibitem{IEEEhowto:kopka}
	Ubeyli, E. D ,\quotes{Decision support systems for time-varying biomedical signals: EEG signals classification},  \emph{Expert Systems with Applications},\hskip 1em plus
  0.5em minus 0.4em\relax Vol. 36, pp. 2275-2284, 2009.
  \bibitem{IEEEhowto:kopka}
Ubeyli, E. D., Guler, I,\quotes{Comparison of eigenvector methods with classical and model-based methods in analysis of internal carotid arterial Doppler signals},  \emph{Computers in Biology and Medicine},\hskip 1em plus
  0.5em minus 0.4em\relax Vol. 33, no.6, pp. 473-493, 2003.
  \bibitem{IEEEhowto:kopka}
		R.B. Pachori,\quotes{Discrimination between ictal and seizure-free EEG signals using empirical mode decomposition},  \emph{Research Letters in Signal Processing 2008},\hskip 1em plus
  0.5em minus 0.4em\relax Article ID 293056, 2008.
  \bibitem{IEEEhowto:kopka}
 R.J. Oweis, E.W. Abdulhay,\quotes{Seizure classification in EEG signals utilizing Hilbert-Huang transform},\emph{Bio Medical Engineering On Line },\hskip 1em plus
  0.5em minus 0.4em\relax December 2011.
   \bibitem{IEEEhowto:kopka}
R.B. Pachori, V. Bajaj,\quotes{Analysis of normal and epileptic seizure EEG signals using empirical mode decomposition},  \emph{Computer Methods and Programs in Biomedicine},\hskip 1em plus
  0.5em minus 0.4em\relax pp. 373-381, December 2011.
  \bibitem{IEEEhowto:kopka}
	V. Bajaj, R.B. Pachori ,\quotes{EEG signal classification using empirical mode decomposition and support vector machine},  \emph{Proceedings International Conference on Soft Computing for Problem Solving},\hskip 1em plus
  0.5em minus 0.4em\relax pp. 623-635, December 2011.
  
  \bibitem{IEEEhowto:kopka}
	S. Li, et al. ,\quotes{Feature extraction and recognition of ictal EEG using EMD and SVM},  \emph{Computers in Biology and Medicine},\hskip 1em plus
  0.5em minus 0.4em\relax pp. 807-816, August 2013.
  \bibitem{IEEEhowto:kopka}
	V. Bajaj, R.B. Pachori,\quotes{Epileptic seizure detection based on the instantaneous area of analytic intrinsic mode functions of EEG signals},  \emph{Biomedical Engineering Letters 3},\hskip 1em plus
  0.5em minus 0.4em\relax pp. 17-21, March 2013.
  \bibitem{IEEEhowto:kopka}
	V. Bajaj, R.B. Pachori,\quotes{Classification of seizure and non-seizure EEG signals using empirical mode decomposition},  \emph{IEEE Transactions on Information Technology in Biomedicine},\hskip 1em plus
  0.5em minus 0.4em\relax pp. 1135-1142, November 2012.
  \bibitem{IEEEhowto:kopka}
	N.Ur. Rehman, Y. Xia, D.P. Mandic ,\quotes{Application of multivariate empirical mode decomposition for seizure detection in EEG signals},  \emph{Proceedings Annual International Conference of IEEE Engineering in Medicine and Biology Society},\hskip 1em plus
  0.5em minus 0.4em\relax August 31, 2010-September 04, 2010.
  \bibitem{IEEEhowto:kopka}
	K. Dragomiretskiy and D. Zosso,\quotes{Variational Mode Decomposition},  \emph{IEEE Transactions on Signal Processing},\hskip 1em plus
  0.5em minus 0.4em\relax vol. 62, no. 3, pp. 531-544, Feb.1 2014.
  \bibitem{IEEEhowto:kopka}
	N. Klügel,\quotes{Practical empirical mode decomposition for audio synthesis},  \emph{Proc. Int. Conf. Digital Audio Effects (DAFx-12)},\hskip 1em plus
  0.5em minus 0.4em\relax pp. 15-18, 2012.
  \bibitem{IEEEhowto:kopka}
	B. Barnhart andW. Eichinger,\quotes{Empirical mode decomposition applied to solar irradiance, global temperature, sunspot number, and CO2 concentration data},  \emph{J. Atmospher. Solar-Terrestrial Phys},\hskip 1em plus
  0.5em minus 0.4em\relax vol. 73, no. 13, pp. 1771-1779, Aug. 2011.
  \bibitem{IEEEhowto:kopka}
	S.Assous,A. Humeau, and J.-P. L?huillier,\quotes{Empirical mode decomposition applied to laser Doppler flowmetry signals: Diagnosis approach},  \emph{Proc. IEEE Eng. Med. Biol. Conf. (EMBC)},\hskip 1em plus
  0.5em minus 0.4em\relax vol. 2, pp. 1232-1235, January 2005.
  \bibitem{IEEEhowto:kopka}
	A. O. Andrade, S. Nasuto, P. Kyberd, C. M. Sweeney-Reed, and F.Van Kanijn,\quotes{EMG signal filtering based on empirical mode decomposition},  \emph{Biomed. Signal Process. Control},\hskip 1em plus
  0.5em minus 0.4em\relax vol. 1, no. 1, pp. 44-55, January 2006
  \bibitem{IEEEhowto:kopka}
S. Liu, Q. He, R. X. Gao, and P. Freedson,\quotes{Empirical mode decomposition applied to tissue artifact removal from respiratory signal},  \emph{Proc. IEEE Eng. Med. Biol. Conf. (EMBC)},\hskip 1em plus
  0.5em minus 0.4em\relax pp. 3624-3627 January 2008.
  \bibitem{IEEEhowto:kopka}
	I. Mostafanezhad, O. Boric-Lubecke, V. Lubecke, D. P. Mandic,\quotes{Application of empirical mode decomposition in removing fidgeting interference in Doppler radar life signs monitoring devices},  \emph{Proc .IEEE Eng. Med. Biol. Conf. (EMBC)},\hskip 1em plus
  0.5em minus 0.4em\relax pp. 340-343 January  2009.
  \bibitem{IEEEhowto:kopka}
	G. Rilling, P. Flandrin, and P. Gonçalvès ,\quotes{On empirical mode decomposition and its algorithms},  \emph{Proc. IEEE-EURASIP Workshop Nonlinear Signal Image Process. (NSIP)},\hskip 1em plus
  0.5em minus 0.4em\relax vol. 3, pp. 8-11, 2003.
  \bibitem{IEEEhowto:kopka}
R.G. Andrzejak, et al,\quotes{Indications of nonlinear deterministic and finite-dimensional structures in time series of brain electrical activity: dependence on recording region and brain state},  \emph{Physical Review E},\hskip 1em plus
  0.5em minus 0.4em\relax vol. 64, Article ID 061907, 2001.
  \bibitem{IEEEhowto:kopka}
		M.E. Cohen, D.L. Hudson, P.C. Deedwania,\quotes{Applying continuous chaotic modeling to cardiac signal analysis},  \emph{IEEE Engineering in Medicine and Biology Magazine },\hskip 1em plus
  0.5em minus 0.4em\relax pp. 97-102, October 1996.
  \bibitem{IEEEhowto:kopka}
R.B Pachori, S Patidar,\quotes{Epileptic seizure classification in EEG signal using second order difference plot of intrinsic mode functions},  \emph{Computer Programs and Methods in Bio-medicine},\hskip 1em plus
  0.5em minus 0.4em\relax pp. 494-502, 2014.
  \bibitem{IEEEhowto:kopka}
	J. Fell, J. Röschke, K. Mann, C. Schäffner,\quotes{Discrimination of sleep stages: A comparison between spectral and nonlinear EEG measures},  \emph{Electroencephalogr. Clin. Neurophysiol},\hskip 1em plus
  0.5em minus 0.4em\relax pp. 401-410, 1996.
  \bibitem{IEEEhowto:kopka}
	R.B Pachori, R Sharma, U Acharya ,\quotes{Application of Entropy measures on Intrinsic Mode functions for the automated identification of Focal EEG signals},  \emph{Entropy},\hskip 1em plus
  0.5em minus 0.4em\relax pp. 669-691, 2015.
  
 

\end{thebibliography}
%

%

\begin{IEEEbiography}[{\includegraphics[width=1in,height=1.25in,clip,keepaspectratio]{picture}}]{John Doe}
\blindtext
\end{IEEEbiography}




\end{document}